\documentclass[12pt]{article}
\usepackage{latexsym}
\usepackage{amsmath}
\usepackage{amsfonts}
\usepackage{amssymb}
\usepackage{amscd}
\usepackage{fancybox}
\usepackage{cite}
\usepackage{amsmath,amsfonts,amsbsy}
\usepackage{pstricks,pst-node}
\usepackage[small,bf,hang]{caption2}

\usepackage{float}

\psset{unit=1.3cm,linewidth=.5pt,radius=.2}  

\usepackage{enumitem}                       
\usepackage{multirow}                     
\usepackage{float}                          
\usepackage{lscape}                         
\usepackage{bm}

\def\hybrid{\topmargin -20pt    \oddsidemargin 0pt
        \headheight 0pt \headsep 0pt
        \textwidth 6.25in       
        \textheight 9.5in       
        \marginparwidth .875in
        \parskip 5pt plus 1pt   \jot = 1.5ex}

\hybrid

\newcommand{\cK}{{\cal K}}
\newcommand{\cL}{{\cal L}}
\newcommand{\cM}{{\cal M}}
\newcommand{\cN}{{\cal N}}

\newcommand{\cP}{{\cal P}}

\newcommand{\cS}{{\cal S}}

\newcommand{\beq}{\begin{equation}}
\newcommand{\eeq}{\end{equation}}
\newcommand{\bi}{\begin{itemize}}
\newcommand{\ei}{\end{itemize}}
\newcommand{\bea}{\begin{eqnarray}}
\newcommand{\eea}{\end{eqnarray}}
\newcommand{\ba}{\begin{array}}
\newcommand{\ea}{\end{array}}
\newcommand{\bt}{\begin{tabular}}
\newcommand{\et}{\end{tabular}}
\newcommand{\bc}{\begin{center}}
\newcommand{\ec}{\end{center}}

\def\theequation{\arabic{section}.\arabic{equation}}

\newcommand{\ket}[1]{|#1\rangle}

\newcommand{\ft}[2]{{\textstyle {\frac{#1}{#2}} }}

\begin{document}

\begin{titlepage}
\begin{center}

\hfill UG-08-12 \\

\vskip 1.5cm



{\Large \bf A Topologically Massive Gauge Theory\\ with 32
Supercharges
\\[0.2cm]}

\vskip 1.5cm

{\bf Eric A.~Bergshoeff\, and Olaf Hohm} \\

\vskip 25pt

{\em Centre for Theoretical Physics, University of Groningen, \\
Nijenborgh 4, 9747 AG Groningen, The Netherlands \vskip 5pt }

{email: {\tt E.A.Bergshoeff@rug.nl, O.Hohm@rug.nl}} \\

\vskip 0.8cm

\end{center}

\vskip 2cm

\begin{center} {\bf ABSTRACT}\\[3ex]

\begin{minipage}{13cm}
We construct a novel topologically massive abelian Chern-Simons
gauge theory with 32 global supersymmetries in three spacetime
dimensions. In spite of the 32 supercharges, the theory exhibits
massive excitations only up to spin 1. The possibility of such a
multiplet shortening is due to the presence of non-central
R-symmetry generators in the Poincar\'e superalgebra, whose
supermultiplets are determined.

\end{minipage}

\end{center}

\noindent

\vfill

October 2008

\end{titlepage}

\section{Introduction}\setcounter{equation}{0}
According to standard folklore supersymmetric field theories are
restricted to 16 supercharges in case of global supersymmetry or to
32 supercharges in case of local supersymmetry. This `no-go theorem'
follows from the requirement that the states of a supermultiplet
should not exceed spin 1 (without gravity) or spin 2 (with gravity).
This conclusion applies to four dimensions, where a notion of spin
can be readily defined, but also to all higher dimensions $4 < D \le
11$, which are related to the four-dimensional case via dimensional
reduction.

However, in three dimensions the situation is more subtle. First of
all, in the massless case there is no notion of spin or helicity, as
the massless little group degenerates to the trivial $\text{SO}(1)$.
Moreover, scalars are dual to vectors -- obscuring the difference
between massless scalar and vector multiplets --, while states of
`higher spin' are topological. Therefore, supermultiplets might
exist for any number ${\cal N}$ of supersymmetries. Indeed, free
globally supersymmetric theories possessing only massless scalars
and Majorana spinor fields can be written for any ${\cal N}$
\cite{Marcus:1983hb}. These theories seem, however, not to be
extendable to non-linear theories, at least not in the form of
non-linear $\sigma$--models \cite{deWit:1992up}.

Notwithstanding the degenerate massless case, a notion of spin does
exist in the massive case, where the little group becomes $\text{SO}(2)$.
Thus, here one expects a priori the same bounds as in the massless case
in $D=4$. However, it turns
out that the three-dimensional Poincar\'e superalgebra allows an extension
by non-central R-symmetry generators, which does not have an
analogue in higher dimensions \cite{Nahm:1977tg}. For ${\cal N}=8$ this non-standard
superalgebra appears as the super-isometry algebra of the IIB
plane wave background \cite{Blau:2001ne} and has recently re-appeared
in the study of mass-deformed
multiple M2-branes \cite{Gomis:2008cv,Hosomichi:2008qk}.  In
somewhat different manifestations the same algebra also occurs in the context of intersecting five-branes
\cite{Itzhaki:2005tu} and in certain sectors of the
$AdS_5/CFT_4$ correspondence
\cite{Lin:2005nh,Hofman:2006xt,Beisert:2005tm}.
In
this paper we will study the supermultiplets and Poincar\'e
invariant field theories based on these unconventional superalgebras. One finds an
unexpected type of multiplet shortening, which allows to increase
the number of supercharges beyond the barrier mentioned above. As
the main result of this paper, we derive a globally supersymmetric
massive ${\mathcal N=16}$ theory, which exhibits 32 supercharges in
spite of the fact that the maximum spin is 1. Specifically, this is
an abelian gauge theory, in which the vectors become topologically
massive due to the presence of a Chern-Simons term
\cite{Deser:1981wh,Schonfeld:1980kb,Townsend:1983xs,Deser:1984kw}.
To derive this model we will follow a method which has recently
\cite{Bergshoeff:2008cz,Bergshoeff:2008ix,Bergshoeff:2008bh} been
pursued in order to derive the ${\cal N}=8$ membrane actions of
\cite{Bagger:2007jr} from the corresponding supergravity theories
\cite{Nicolai:2001ac,deWit:2003ja}. Applying the same technique to
maximal ${\cal N}=16$ supergravity, one finds, surprisingly, that in
contrast to higher dimensions, the topological supergravity
fields can be decoupled, leaving a non-trivial ${\cal
N}=16$ matter theory. By starting from ungauged supergravity we
recover the free massless theories of \cite{Marcus:1983hb}, while the
massive theory is obtained by starting from  gauged supergravity
\cite{Nicolai:2000sc,Nicolai:2001sv}.

The organization of the paper is as follows. In sec.~2 we determine
the massive supermultiplets in presence of the non-central
R-symmetry charges. In sec.~3 we review the Poincar\'e invariant
field theories for massive scalar multiplets with ordinary central
charges (${\cal N}=2$) and with non-central charges ($\cN=4$ and
$\cN=8$). In sec.~4 we consider vector multiplets and determine the
${\cal N}=16$ topologically massive gauge theory. We conclude with
an outlook in sec.~5. Our conventions are summarized in an appendix.

\section{Massless and massive supermultiplets}
In this section we determine the massive supermultiplets for the
Poincar\'e superalgebra with non-central charges. For completeness
we first review the standard massless and massive multiplets as
well as ordinary BPS multiplets. The reader only interested in the
multiplet shortening due to the non-central charges might skip this part and proceed
directly to sec.~\ref{noncentral}

\subsection{Standard Poincar\'e superalgebra}\label{standard}
The standard ${\cal N}$-extended Poincar\'e superalgebra for
Majorana supercharges $Q_{\alpha}^{i}$ reads
 \bea\label{Poincare}
   \{ Q_{\alpha}^{i},Q_{\beta}^{j} \} \ = \
  2\left(\gamma^{\mu}C\right)_{\alpha\beta}P_{\mu}\delta^{ij}
  \;,
 \eea
where $i,j,\ldots = 1,\ldots,{\cal N}$ and $C_{\alpha\beta}$ is the
charge-conjugation matrix. For our $\text{SO}(1,2)$ spinor conventions we refer to
appendix \ref{3Dspinor}. The other commutation relations are
standard, expressing the Poincar\'e algebra and the transformation
properties of the supercharges under the Lorentz group.

We start with the supermultiplets of (\ref{Poincare}) in the
massless case $P^{2}=0$, which has been analyzed in
\cite{deWit:1992up}. For $P_{\mu}=(\omega,0,\omega)$ the
superalgebra reads
 \bea\label{masslesssusy}
  \{ Q_{\alpha}^{i},Q_{\beta}^{j}\} \ = \ 4\omega
  \left(\begin{array}{cc} 0 & 0 \\ 0 & 1
  \end{array}\right)_{\alpha\beta}\delta^{ij}\;.
 \eea
As usual, half of the supercharges disappear, leaving a Clifford
algebra for $\text{SO}({\cal N})$, spanned by $Q_{2}^{i}$. In
addition, there is a fermion number operator $F$, which anticommutes
with the supercharges and satisfies $F^2={\bf 1}$. This extends the
algebra to the Clifford algebra of $\text{SO}({\cal N}+1)$. Since in
the massless case there is no notion of spin, the only thing we can
consider when analyzing the multiplets is the number of bosonic and
fermionic states, respectively. These are given by (half of) the
dimension of the Clifford algebra, which are known for all values of
${\cal N}+1$.
The result is
summarized in tab.~1.

We now turn to the massive case $P^{2}=M^2$. We boost into the rest
frame, $P_{\mu}=(M,0,0)$, and redefine the supercharges according to
\cite{Hohm:2004rc}
 \bea
  (a^{i})^{\dagger} \ = \ \ft12\left(Q_{1}^{i}+iQ_{2}^{i}\right)\;,
  \qquad
  a^{i} \ = \ \ft12\left(Q_1^{i}-iQ_{2}^{i}\right)\;.
 \eea
The superalgebra (\ref{Poincare}) reads in this basis
 \bea\label{massiveosc}
 \begin{split}
  \{ a^{i},(a^j)^{\dagger}\} \ &= \
  M\delta^{ij}\;, \\
  \{ a^{i},a^j \} \ &= \ \{ (a^{i})^{\dagger},(a^{j})^{\dagger}\} \ =
  \ 0\;.
 \end{split}
 \eea
This redefinition is such that the supercharges and their conjugates
can be interpreted as lowering and raising operators. Moreover, they
increase and decrease the space-time helicity with respect to the
little group $\text{SO}(2)$ \cite{Hohm:2004rc}. In the massive case
the supermultiplets are therefore standard, carrying $2^{\cal N}$
states with helicities ranging from $j$ to $j+{\cal N}/2$. In
particular, for ${\cal N}=8$ the helicities should go up to 2 and
therefore there can be no massive scalar multiplets based on the
ordinary superalgebra. This is in contrast to the non-central
charges to be analyzed below.
\begin{table}[tdd]\label{massless}
  \begin{center}
   $\begin{array}{|c|c|c|c|c|c|c|c|c|c|   }
\hline
{\cal N}   &  1 &  2 & 3 & 4 & 5 &  6 &   7 & 8 &n+8\\
\hline
d_n &  1 &  2 & 4 & 4 &  8 & 8 & 8 & 8 &16d_n\\
   \hline
   \end{array}$
   \caption{Number of bosonic states $d_n$ for massless ${\cal N}$--extended supermultiplets}
  \end{center}
 \end{table}

\subsection{Centrally extended Poincar\'e superalgebra}
As an example of a centrally extended superalgebra we consider
the ${\cal N}=2$ super-Poincar\'e algebra
 \bea\label{N2ext}
   \{ Q_{\alpha}^{i},Q_{\beta}^{j} \} \ = \
  2\left(\gamma^{\mu}C\right)_{\alpha\beta}P_{\mu}\delta^{ij}
  +2mi\varepsilon^{ij}C_{\alpha\beta}(Z_1-Z_2)\;.
 \eea
Note that here we have introduced two $\text{U}(1)$ generators,
$Z_1$ and $Z_2$, whose commutation relations with the supercharges
read
 \bea
  [Z_1,Q_{\alpha}^{i}] \ = \ \varepsilon^{ij}Q^{j}_{\alpha}\;,
  \qquad
  [Z_2,Q^{i}_{\alpha}] \ = \ \varepsilon^{ij}Q^{j}_{\alpha}\;.
 \eea
This implies that the combination $Z\equiv Z_1-Z_2$ appearing on the
right-hand side of the superalgebra (\ref{N2ext}) commutes with the
supercharges and therefore represents a central extension. This is
also required by consistency with the super-Jacobi identities. The
combination $R\equiv Z_1+Z_2$, on the other hand, rotates the
supercharges according to their $\text{SO}(2)$ indices and thus
represents the $\text{SO}(2)\cong \text{U}(1)$ R-symmetry. Due to
the central charges, BPS multiplets are possible in the massive
case. First, the oscillator algebra (\ref{massiveosc}) gets replaced
by
 \bea
   \{ a^{i},(a^j)^{\dagger}\} \ = \ M\delta^{ij}
   +im\varepsilon^{ij}Z\;.
 \eea
The eigenvalues of the matrix appearing on the right-hand side are
given by $M\pm m|Z|$. Unitarity implies therefore the following
bound
 \bea
  M\geq m|Z| \;.
 \eea
In case this bound is saturated, one of the oscillators trivializes
and thus the multiplets are as for ${\cal N}=1$. For instance, a
scalar multiplet contains spins $(0,\ft12)$. Since this carries only
real degrees of freedom, it cannot transform under the required
$\text{U}(1)$ R-symmetry. Thus, we have to complexify, leading to the
${\cal N}=2$ multiplet $(0,\ft12)\oplus (0,\ft12)$. We observe that
parity-odd multiplets are natural in the presence of central charges.

\subsection{Non-centrally extended Poincar\'e
superalgebra}\label{noncentral} The possibility of a non-centrally
extended superalgebra arises for ${\mathcal N}\ge 4$. For ${\cal
N}=4$, the super-Poincar\'e algebra can be extended by the following
non-central charges:
 \bea\label{susyalg}
  \{ Q_{\alpha}^{i},Q_{\beta}^{j} \} \ = \
  2\left(\gamma^{\mu}C\right)_{\alpha\beta}P_{\mu}\delta^{ij}
  +2mC_{\alpha\beta}\varepsilon^{ijkl}M_{kl}\;,
 \eea
where $M_{ij}$ denote the $\text{SO}(4)$ R-symmetry generators.  In
particular, they do not commute, but instead satisfy the standard
relations
 \bea\label{so4}
  \begin{split}
   [M_{ij},M_{kl}] \ &= \
   -2\left(\delta_{k[i}M_{j]l}-\delta_{l[i}M_{j]k}\right)\;, \\
   [M^{ij},Q_{\alpha}^{k}] \ &= \ 2\delta^{k[i}Q_{\alpha}^{j]}\;.
  \end{split}
 \eea
Algebras of this type also appear in the context of AdS supergroups,
where the supercharges generically close into the R-symmetry group.
The peculiar property here, however, is that this represents a
consistent algebra for Poincar\'e supersymmetry, i.e., despite the
commuting translations, the particular choice (\ref{susyalg})
containing an $\text{SO}(4)$ Levi-Civita symbol satisfies the
super-Jacobi identities \cite{Lin:2005nh}. This non-central
extension is also possible for ${\cal N}>4$. We will mainly consider
the case of ${\cal N}$ being $k$ multiples of $4$, for which one has
$k$ copies of the algebra (\ref{susyalg}).\footnote{We should note,
however, that this is not a direct sum, since there is only a single
energy-momentum operator $P_{\mu}$.} In this case, the ${\rm
SO}({\cal N})$ R-symmetry group will be broken to
$\text{SO}(4)^k$.\footnote{Theories with ${\cal N}=6$, breaking the
R-symmetry to $SO(4)\times U(1)$, and $\cN=5$ were considered in
\cite{Hosomichi:2008jb,Gomis:2008vc}.}

Let us now turn to the supermultiplets of (\ref{susyalg}). We note
that (\ref{susyalg}) is related to a central extension of the
superalgebra $\frak{su}(2|2)$, which appeared in the $AdS_5/CFT_4$
correspondence. More precisely, the central charges in that algebra
can be reinterpreted as a $2+1$ dimensional energy-momentum operator
$P_{\mu}$, while the Lorentz generators do not appear, but rather
represent outer automorphisms \cite{Beisert:2005tm,Hofman:2006xt}.
The representation theory of the latter algebra has been developed
in \cite{Beisert:2006qh}. Here, we are going to apply the standard
little group technique to (\ref{susyalg}) and determine the
supermultiplets via introducing oscillators.

In case of $m\neq 0$ there are no massless representations of the
superalgebra (\ref{susyalg}), which can be easily seen as follows
\cite{Lin:2005nh}. For $P_{\mu}=(\omega,0,\omega)$ the algebra reads
 \bea
  \{ Q_{\alpha}^{i},Q_{\beta}^{j}\} \ = \ 4\omega
  \left(\begin{array}{cc} 0 & 0 \\ 0 & 1
  \end{array}\right)_{\alpha\beta}\delta^{ij}
  -2im\left(\begin{array}{cc} 0 & 1 \\ -1 & 0
  \end{array}\right)_{\alpha\beta}\varepsilon^{ijkl}M_{kl}\;.
 \eea
Thus, like in eq.~(\ref{masslesssusy}), $\{Q_{1}^{i},Q_{1}^{j}\}$
vanishes, and so in a positive-definite Hilbert space $Q_{1}^{i}$
has to act trivially. On the other hand, the off-diagonal bracket
$\{Q_{1}^{1},Q_{2}^{2}\}$, for instance, is proportional to
$M^{34}$, and therefore also this $\text{SO}(4)$ generator has to act
trivially. However, this is in conflict with the fact that according
to (\ref{so4}) the supercharges change the $\text{SO}(4)$ quantum numbers,
and so the states are generically not singlets. Thus, massless
representations can only exist for $m=0$.

We next consider the massive case. The oscillator algebra now reads
 \bea\label{Noncenos}
  \{ a^{i},(a^j)^{\dagger}\} \ = \
  M\delta^{ij}+m\varepsilon^{ijkl}M_{kl}\;.
 \eea
It turns out to be convenient to construct the representations using
$\text{SU}(2)$ spinor indices via the isomorphism $\text{SO}(4)\cong
\text{SU}(2)_{L}\times \text{SU}(2)_{R}$. Specifically, the oscillators are
bispinors
\begin{equation}
a_{a\dot{a}}=\Gamma^{i}_{a\dot{a}}\,a^{i}\,,
\end{equation}
where $\Gamma^{i}_{a\dot{a}}$ are $\text{SO}(4)$ gamma matrices and we use
undotted and dotted indices for $\text{SU}(2)_{L}$ and $\text{SU}(2)_{R}$,
respectively. The $\text{SO}(4)$ generators decompose accordingly into the
symmetric $\text{SU}(2)_{L,R}$ generators $M^{ab}$ and
$M^{\dot{a}\dot{b}}$. For further details on this notation we refer
to  appendix \ref{so4app}. Using this notation the algebra
(\ref{Noncenos}) reads
 \bea\label{superspinor}
 \begin{split}
  \{ a_{a\dot{a}},a_{b\dot{b}}^{\dagger} \} \ &= \
  -2M\varepsilon_{ab}\varepsilon_{\dot{a}\dot{b}}
  -4m\left(\varepsilon_{\dot{a}\dot{b}}M_{ab}-\varepsilon_{ab}M_{\dot{a}\dot{b}}\right)\;,
  \\
  \{ a_{a\dot{a}},a_{b\dot{b}} \} \ &= \ \{
  a_{a\dot{a}}^{\dagger},a_{b\dot{b}}^{\dagger} \} \ = \ 0\;.
 \end{split}
 \eea
We note that the two $\text{SU}(2)$ factors enter with a relative
minus sign, which is due to their respective self-duality and
anti-self-duality, cf.~eq.~(\ref{relsign}) in the Appendix. In
addition, the supercharges satisfy the commutation relations
(\ref{semiraising}), indicating that they act as raising and
lowering operators for the $\text{SU}(2)$ quantum numbers. To be
more precise, if one writes the spinor indices as $a=(+,-)$ and
$\dot{a}=(\dot{+},\dot{-})$, then an undotted or dotted `+' index
indicates that the $\text{SU}(2)_{L,R}$ spin quantum number is
increased by $\ft12$, while a `--' index indicates that it is
decreased by $\ft12$. Moreover, $M_{+-}$ corresponds to the $J_3$
operator and thus measures the spin quantum number $\ell$, while
$M_{++}$ and $M_{--}$ are $SU(2)$ raising and lowering operators.

In order to construct shortened supermultiplets we must impose a
generalized BPS condition. To see how this works, let us consider the bracket
 \bea
  \{ a_{+\dot{+}},(a_{+\dot{+}})^{\dagger}\} \ = \ -\{a_{+\dot{+}},a_{-\dot{-}}^{\dagger}\}
  \ = \  2M-4m(J_3^{L}-J_3^{R})\;,
 \eea
where we used
$(a_{+\dot{+}})^{\dagger}=-\varepsilon^{+-}\varepsilon^{\dot{+}\dot{-}}a^{\dagger}_{
-\dot -}$. In case the BPS-like condition $M =
2m(\ell_{L}-\ell_{R})$ is satisfied, positivity of the Hilbert space
 implies that $a^{\dagger}_{-\dot{-}}$ is deactivated. Similarly,
 one derives from (\ref{superspinor}) that each of the four possible
raising operators is deactivated provided the corresponding BPS
condition is satisfied:
 \bea\label{BPScond}
  \begin{split}
   &a^{\dagger}_{+\dot{+}}\;: \qquad M=-2m(\ell_{L}-\ell_{R})\;, \\
   &a^{\dagger}_{+\dot{-}}\;: \qquad M=-2m(\ell_{L}+\ell_{R})\;, \\
   &a^{\dagger}_{-\dot{+}}\;: \qquad M=2m(\ell_{L}+\ell_{R})\;, \\
   &a^{\dagger}_{-\dot{-}}\;: \qquad M=2m(\ell_{L}-\ell_{R})\;.
  \end{split}
 \eea
Note that, in contrast to ordinary BPS multiplets,
different sets of supercharges become trivial, depending on which
states they act.

Let us now turn to the construction of the supermultiplets for
${\cal N}=4$, which is the first non-trivial case. We label the
states $\ket{j;\ell_{L},\ell_{R}}$ by the space-time helicity $j$
and, in the second and third entry, by spin quantum numbers
$\ell_{L}$ and $\ell_{R}$ of $\text{SU}(2)_{L}$ and $\text{SU}(2)_{R}$,
respectively. As usual, we start from a `Clifford vacuum' as the
lowest state. For the smallest multiplets we choose
 \bea
  \ket{\Omega} \ = \ \ket{j_0;0,-\ft12}\;,
 \eea
which is annihilated by all $a_{a\dot{b}}$.  Assuming $M=m$,
(\ref{BPScond}) implies that only $a_{-\dot{+}}^{\dagger}$ and
$a_{+\dot{+}}^{\dagger}$ are active. Thus we obtain two states with
helicity $j_0+\ft12$\,: $\ket{j_0+\ft12;\ft12,0}$ and
$\ket{j_0+\ft12;-\ft12,0}$. It is not possible to act a second time
with the creation operators, which is the main reason for the
multiplet shortening. To see this we note that due to
(\ref{BPScond}) and the anticommutativity of the oscillators on,
say, $\ket{j_0+\ft12;\ft12,0}$ only $a_{+\dot{-}}^{\dagger}$ can be
potentially non-zero. However, one finds
 \bea
   a^{\dagger}_{+\dot{-}}\ket{j_0+\ft12;\ft12,0} \ = \
   a^{\dagger}_{+\dot{-}}a^{\dagger}_{+\dot{+}}\ket{\Omega}
   \ = \ -a^{\dagger}_{+\dot{+}}a^{\dagger}_{+\dot{-}}\ket{\Omega}
   \ = \ 0\;,
 \eea
where we used in the last equation that $a^{\dagger}_{+\dot{-}}$ is
inactive on the vacuum $\ket{\Omega}$. Similarly, one derives that
there are no other states of helicity higher than $j_0+\ft12$.
Finally, by acting with the $\text{SU}(2)$ raising and lowering
operators $M_{++}$, etc., the states combine into complete
$\text{SU}(2)$ representations. If we choose $j_0=-\ft12$, this
${\cal N}=4$ multiplet consists of two complex scalars $\phi_{a}$
and two Dirac fermions $\chi_{\dot{a}}$, corresponding to the
following states:
 \bea\label{N=4multiplet}
  \begin{CD}
   \chi_{\dot{-}}\ = \ \ket{-\ft12;0,-\ft12}
   @>\;\;a_{+\dot{+}}^{\dagger}\;\;>>
   \phi_{+} \ = \ \ket{0;\ft12,0}\\
   @VM_{\dot{+}\dot{+}}VV @VVM_{--}V\\
   \chi_{\dot{+}} \ = \ \ket{-\ft12;0,\ft12} @>>a_{-\dot{-}}^{\dagger}>
   \phi_{-} \ = \ \ket{0;-\ft12,0}
 \end{CD}
\eea The action of those operators not indicated in
(\ref{N=4multiplet}) is either trivial or equivalent to the
consecutive action of the given ones. For instance,
$a^{\dagger}_{-\dot{+}}$ acting on the vacuum leads to the same
state as $M_{\dot{+}\dot{+}}$ and $a_{-\dot{-}}^{\dagger}$, which
follows from the commutation relations (\ref{semiraising}),
 \bea
  a_{-\dot{-}}^{\dagger}M_{\dot{+}\dot{+}}\ket{\Omega} \ = \
  -[M_{\dot{+}\dot{+}},a_{-\dot{-}}^{\dagger}]\ket{\Omega} \ = \
  -a^{\dagger}_{-\dot{+}}\ket{\Omega}\;.
 \eea
We note that there is no state with space-time helicity $+\ft12$ and
therefore the multiplets are parity-odd. Summarizing, the action of
the creation operators on the vacuum raises the helicity from $j_0$
to $j_0+\ft12$ and converts the complex $\text{SU}(2)_R$
representation $2_{\mathbb C} = (0, \ft12)$ of the ground state into
a $\text{SU}(2)_L$ representation ${\bar 2}_{\mathbb{C}} =
(\ft12,0)$ of the next $j_0+\ft12$ state. This pattern will repeat
itself in the ${\cal N}>4$ cases which we will discuss now.

First, we discuss ${\cal N}=8$, for which the non-central charges
break the R-symmetry group to
 \bea\label{SO(4)2}
  \text{SO}(8)\ \rightarrow \ \text{SO}(4)^{+}\times \text{SO}(4)^{-}\ \cong \
  \text{SU}(2)_L^{+}\times \text{SU}(2)_{R}^{+}\times \text{SU}(2)^{-}_{L}\times
  \text{SU}(2)_{R}^{-}\;.
 \eea
In this case one has two copies of (\ref{superspinor}), say, with
raising operators $a^{\dagger}_{a\dot{a}}$ and
$b^{\dagger}_{a\dot{a}}$, where we do not distinguish between
$\text{SU}(2)$ indices on the different operators, but simply
understand that $a^{\dagger}$ acts on the $\text{SU}(2)$ factors of
$\text{SO}(4)^{+}$ and $b^{\dagger}$ on those of $\text{SO}(4)^{-}$.
Analogous to ${\cal N}=4$ we take the vacuum to be
 \bea\label{N=8vacuum}
   \ket{\Omega} \ = \ \ket{j_0;0,-\ft12,0,-\ft12}\;,
 \eea
where the labels
$(\ell_{L}^{+},\ell_{R}^{+},\ell_{L}^{-},\ell_{R}^{-})$ refer to the
helicity quantum numbers of (\ref{SO(4)2}). In the following we will
assume that the factors $m$ are the same for both $\text{SO}(4)$
sectors, such that the BPS condition $M=m$ gives rise to the maximal
possible multiplet shortening. Other cases of less multiplet
shortening can be analyzed along similar lines. On (\ref{N=8vacuum})
one can then act either with the $a^{\dagger}$ or the $b^{\dagger}$
operators, which in analogy to ${\cal N}=4$ gives four types of
states, with space-time helicity $j_0+\ft12$. In contrast to ${\cal
N}=4$ it is now possible to act a second time with raising
operators, giving states of the type
$a^{\dagger}b^{\dagger}\ket{\Omega}$. Thus, the multiplet contains
also states with space-time helicity $j_0+1$. In particular, in case
of $j_0=-\ft12$, which we will consider in the following, the
multiplet is parity-even, consisting of $8$ bosons and $8$ fermions.
Using the following decomposition of the $\text{SO}(8)$
representations ${\bf 8}_{V}$, ${\bf 8}_{S}$ and ${\bf 8}_{C}$ under
(\ref{SO(4)2})
 \bea
  \begin{split}
   {\bf 8}_{V}\ &\rightarrow \
   (\ft12,\ft12,0,0)+(0,0,\ft12,\ft12)\;, \\
   {\bf 8}_{S}\ &\rightarrow \
   (\ft12,0,0,\ft12)+(0,\ft12,\ft12,0)\;, \\
   {\bf 8}_{C}\ &\rightarrow \
   (0,\ft12,0,\ft12)+(\ft12,0,\ft12,0)\;,
  \end{split}
 \eea
one finds that the representations of the supermultiplet are such
that the scalars can be combined into ${\bf 8}_{S}$ ($\phi^{A}$) and
the Majorana spinors into ${\bf 8}_{C}$
($\chi^{\dot{A}}$).\footnote{Here  $I,J,\ldots =1,\ldots,8$,
$A,B,\ldots =1,\ldots,8$ and $\dot{A},\dot{B},\ldots =1,\ldots,8$
denote vector, spinor and conjugate spinor indices of
$\text{SO}(8)$.}

The construction of the multiplets for arbitrary ${\cal N}=4k$
proceeds in exact analogy, using $k$ sets of oscillators. For
instance, in case of ${\cal N}=16$, which will be of relevance
below, the R-symmetry group is broken according to
 \bea\label{SO(16)8}
  \text{SO}(16)\rightarrow \text{SO}(8)\times \text{SO}(8)\;,
 \eea
and then each $\text{SO}(8)$ further according to (\ref{SO(4)2}). The basic
(real) representations decompose under (\ref{SO(16)8}) into
 \bea\label{SO16split}
  \begin{split}
   {\bf 16}_{V}\ &\rightarrow \ ({\bf 8}_{V},{\bf 1})+({\bf 1},{\bf
   8}_{V})\;, \\
   {\bf 128}_{S} \ &\rightarrow \ ({\bf 8}_{S},{\bf 8}_{S})+({\bf
   8}_{C},{\bf 8}_{C})\;, \\
   {\bf 128}_{C} \ &\rightarrow \ ({\bf 8}_{S},{\bf 8}_{C})+({\bf
   8}_{C},{\bf 8}_{S})\;.
  \end{split}
 \eea
The ${\cal N}=16$ superalgebra is spanned by four types of
oscillators, say, $a^{\dagger}$, $b^{\dagger}$, $c^{\dagger}$ and
$d^{\dagger}$. Starting from a vacuum $\ket{\Omega}$ with space-time
helicity $j_0$, one can now create a state
$a^{\dagger}b^{\dagger}c^{\dagger}d^{\dagger}\ket{\Omega}$, which
has helicity $j_0+2$. Thus, starting from helicity $-1$, one obtains
helicity $+1$ and so the multiplet can be parity-even. However, in
spite of the fact that we are dealing with $32$ supercharges, spin-2
states are not required. We finally note that according to
(\ref{SO16split}) the bosonic degrees of freedom can be combined
into the $\text{SO}(16)$ representation ${\bf 128}_{S}$ ($\phi^{A}$)
and the fermionic degrees of freedom into ${\bf 128}_{C}$
($\chi^{\dot{A}}$), in which, however, 32 of the scalars correspond
to St\"uckelberg fields, that will be eaten by the vectors.

\vskip .3truecm

\renewcommand{\arraystretch}{1.5}
\begin{table}[ht]
\begin{center}
$
\begin{array}{||c||c|c|c|c||}
\hline \text{helicity} &  \cN=4 & \cN=8 & \cN=12 &\cN=16 \\ \hline
\hline
j_0 & 2_{\mathbb{C}} & (2,2) & (2,2,2)_{\mathbb{C}} &(2,2,2,2)\\[1mm] \hline
j_0+\ft{1}{2} & {\bar 2}_{\mathbb{C}}& (\bar 2,2)+(2,\bar 2) & (\bar 2,2,2)_{\mathbb{C}}+ 2\ \text{more}&(\bar 2,2,2,2) + 3\ \text{more} \\[1mm]\hline
j_0+1& & (\bar 2,\bar 2) & (\bar 2,\bar 2,2)_{\mathbb{C}}+ 2\ \text{more}& (\bar 2,\bar 2,2,2)+ 5\ \text{more}\\[1mm]\hline
j_0+\ft{3}{2} &  &  & (\bar 2, \bar 2,\bar 2)_{\mathbb{C}} &(\bar 2,\bar 2,\bar 2,2)+ 3\ \text{more}\\[1mm] \hline
j_0+2 &  &  &  &(\bar 2,\bar 2,\bar 2,\bar 2)\\[1mm] \hline\hline
\text{d.o.f.} & 4_B+4_F & 8_B+8_F & 64_B+64_F&128_B+128_F \\[1mm] \hline
\end{array}$
\caption{Multiplet structure for different values of $\mathcal{N}$,
containing the space-time helicity $j$, the (real and complex)
representations of the broken R-symmetry group and the total number
of d.o.f.} \label{tab:multiplets}
\end{center}
\end{table}
\renewcommand{\arraystretch}{1}

To conclude our discussion of the massive representations, we note
that for each ${\cal N}=4k$ the structure is rather similar. One
starts from a ground state $\ket{\Omega}$ which is in the $(2,2,
\ldots , 2)$\ ($k$ factors of 2) representation, being real for $k$
even and complex for $k$ odd. Here, we use the short-hand notation
$2_{\mathbb{C}}=(0,\ft12)$, $\bar{2}_{\mathbb{C}}=(\ft12,0)$,
indicating the spin-$\ft12$ representations of $\text{SU}(2)_L$ and
$\text{SU}(2)_R$, respectively. The first excited state, with
helicity $j_0+\ft12$ is obtained by replacing one of the $2$
representations by a $\bar 2$. This can be done in $k$ different
ways. The next excited state is obtained by replacing in
$\ket{\Omega}$ two $2$ representations by $\bar 2$, which can be
done in ${k\choose 2}$ different ways, etc. We have summarized the
structure of the short multiplets for ${\cal N} = 4,8,12,16$ in
table 2. The first column indicates the space-time helicities (with
$j_0$ the helicity of the ground state), while the other columns
contain the representations under $\bigl(\text{SU}(2)_L\times
\text{SU}(2)_R\bigr)^{\mathcal{N}/4}$.  We note that scalar
multiplets are possible up to ${\cal N}=8$ and spin-1 multiplets up
to ${\cal N}=16$.

\section{Massive scalar multiplets}

In this section we discuss massive supersymmetric field theories for
${\cal N} = 2,4$ and $8$. As the latter is the largest amount of
supersymmetry consistent with maximal spin $\ft12$, we will focus on
massive scalar multiplets. Before doing that, we first briefly
illustrate the notion of parity and helicity, which in three
dimensions is rather different than in four dimensions. Consider a
(real) Majorana fermion $\chi$ of mass $m$. It carries one physical
(propagating) degree of freedom. This means that it cannot carry
both the two helicities $+\ft12$ and $-\ft12$, but only one. Thus,
an action containing only one Majorana fermion necessarily breaks
parity. Consider, for instance, the ${\cal N}=1$ Lagrangian
 \bea\label{N=1action}
   {\cal L} \ = \ \ft12 \partial^{\mu}\phi\partial_{\mu}\phi
   -i\bar{\chi}\gamma^{\mu}\partial_{\mu}\chi-\ft12m^2\phi^2+m\bar{\chi}\chi\;.
 \eea
In order to decide whether this is parity invariant, we have to
define what we mean by parity in three dimensions. In four
dimensions one defines a parity transformation as
$\vec{x}\rightarrow -\vec{x}$, which has determinant $-1$ and thus is
 a reflection. In contrast, in three dimensions this is a
rotation and therefore we should define a parity transformation
rather as inversion $x^{i}\rightarrow -x^{i}$ of one fixed spatial
direction. On spinors this acts as $\chi\rightarrow i\gamma^{i}\chi$, see e.g.~\cite{Bandres:2008vf}.
The fermionic kinetic term in
(\ref{N=1action}) is invariant under this transformation, but the
mass term $m\bar{\chi}\chi$ switches sign. Thus, (\ref{N=1action})
is parity breaking. We note that the sign of the mass term
determines the helicity to be, say, $+\ft12$, and therefore the
corresponding ${\cal N}=1$ supermultiplet has the spin content
$(0,+\ft12)$.

\subsection{${\cal N}=2$} In terms of ${\cal N}=1$ multiplets a priori there are two
possibilities to build ${\cal N}=2$ multiplets: the parity even
$(-\ft12,0,0,\ft12)=(-\ft12,0)\oplus (0,\ft12)$ or the parity odd
$(0,\ft12)\oplus (0,\ft12)$. As we discussed, the standard
Poincar\'e superalgebra admits the former, while the latter is
possible in the presence of central charges. The field content is
given by a complex scalar $\phi$ and a complex (Dirac) fermion
$\chi$. For the $(-\ft12,2\times 0,\ft12)$ multiplet the Lagrangian
is given by
 \bea\label{N=2action}
   {\cal L} \ = \ \ft12\partial^{\mu}\phi^{\star}\partial_{\mu}\phi
   -i\bar{\chi^{\star}}\gamma^{\mu}\partial_{\mu}\chi
   +\ft12 m\left(\bar{\chi}\chi+\bar{\chi^{\star}}\chi^{\star}\right)
   -\ft12 m^{2}\phi^{\star}\phi\;,
  \eea
which is invariant under the ${\cal N}=2$ supersymmetry
transformations
 \bea
   \delta_{\epsilon}\phi \ = \ \bar{\epsilon}\chi\;, \qquad
   \delta_{\epsilon}\chi \ = \
   \tfrac{i}{2}\gamma^{\mu}\partial_{\mu}\phi\; \epsilon^{\star}
   +\ft12 m\phi^{\star}\epsilon\;,
 \eea
parameterized by the complex spinor $\epsilon$. This can also be
obtained through dimensional reduction of the standard ${\cal N}=1$
chiral multiplet in four dimensions. In order to see that
(\ref{N=2action}) has an equal number of positive and negative
helicities and is thus parity invariant, we split the Dirac spinor
into real and imaginary parts, $\chi=\chi _1+i\chi_2$. In terms of
these two Majorana fermions $\chi_{1,2}$, the mass term reads
$m(\bar{\chi_1}\chi_1-\bar{\chi}_2\chi_2)$, i.e., $\chi_1$ and
$\chi_2$ have opposite helicity. The parity transformation leaving
invariant (\ref{N=2action}) is given by $\chi\rightarrow
\gamma^{i}\chi^{\star}$. Since it involves a relative factor of $i$
as compared to the rule employed for the real case, parity exchanges
in addition the real spinors $\chi_1$ and $\chi_2$. This cures the
non-invariance of the ${\cal N}=1$ action under parity by virtue of
the relative sign between the mass terms.

We consider now the parity-odd multiplet. Its action and
supersymmetry rules are given by
 \bea\label{parityoddN=2}
  \begin{split}
   {\cal L} \ = \ &\ft12\partial^{\mu}\phi^{\star}\partial_{\mu}\phi
   -i\bar{\chi^{\star}}\gamma^{\mu}\partial_{\mu}\chi+m\bar{\chi^{\star}}\chi
   -\ft12m^2\phi^{\star}\phi\;, \\
   &\delta\phi \ = \ \bar{\epsilon}\chi\;, \qquad
   \delta\chi \ = \
   \tfrac{i}{2}\gamma^{\mu}\partial_{\mu}\phi\;\epsilon^{\star}
   +\ft12m\phi\epsilon^{\star}\;.
  \end{split}
 \eea
Due to the relative complex conjugation in the mass term, the two
real fermionic fields enter with the same sign for the mass and,
consequently, the action is parity odd. Let us compute the closure
of the supersymmetry algebra,
 \bea\label{N=2central}
   \big[ \delta_{\epsilon_1},\delta_{\epsilon_2}\big] \phi \ = \
   \xi^{\mu}\partial_{\mu}\phi+m\delta_{\Lambda}\phi\;.
 \eea
The algebra closes not only into translations with
 \bea
  \xi^{\mu} \ = \
  \tfrac{i}{2}(\bar{\epsilon}_2\gamma^{\mu}\epsilon_1^{\star}-
  \bar{\epsilon}_1\gamma^{\mu}\epsilon_2^{\star})\;,
 \eea
but also into a $\text{U}(1)$ rotation,
$\delta_{\Lambda}\phi=i\Lambda\phi$, with real parameter
 \bea
   \Lambda \ = \
   -\tfrac{i}{2}(\bar{\epsilon}_2\epsilon_1^{\star}-\bar{\epsilon}_1\epsilon_2^{\star})\;.
 \eea
This is in contrast to the ${\cal N}=1$ and the parity-even ${\cal
N}=2$ theory, where a similar term proportional to $m$ drops out of
the commutator. Note that this $\text{U}(1)$ rotation is not the
$R$--symmetry, as this would rotate the supercharges, violating the
Jacobi identities. Rather, the action actually has a
$\text{U}(1)\times \text{U}(1)$ symmetry, corresponding to the
generators $Z_1$ and $Z_2$ in sec.~2.2, in which the first
$\text{U}(1)$ acts only on the scalar, and the second $\text{U}(1)$
acts only on the spinor. One linear combination of the
$\text{U}(1)$'s corresponds to a central charge
--- appearing on the right-hand side of (\ref{N=2central}) ---,
while the other linear combination corresponds to the R-symmetry and
does not enter the commutator (\ref{N=2central}). The parity-odd
multiplet described by (\ref{parityoddN=2}) is in agreement with the
findings for standard BPS multiplets discussed in sec.~2.2.

\subsection{${\cal N}=4$}
According to the general form of the supermultiplets, for ${\cal
N}=4$ we have four bosonic and four fermionic degrees of freedom.
Since the $\text{SO}(4)$ R-symmetry group is isomorphic to
$\text{SU}(2)_L\times \text{SU}(2)_R$ it is convenient to use
complex notation. The $\cN=4$ multiplets found before consist of two
complex scalars $\phi^a$, transforming under $\text{SU}(2)_L$, and
two complex spinors $\chi_{\dot{a}}$, transforming under
$\text{SU}(2)_R$. This is analogous to the $\text{U}(1)\times
\text{U}(1)$ for $\cN=2$. We use the standard notation that lowering
and raising indices corresponds to complex conjugation,
$(\phi^{a})^{*}=\phi_{a}$, etc. The massive theory is given by the
Lagrangian
 \bea
   {\cal L} \ = \ \ft12 \partial^{\mu}\phi^{a}\partial_{\mu}\phi_a
   -i\bar{\chi}^{\dot{a}}\gamma^{\mu}\partial_{\mu}\chi_{\dot{a}}
   -\ft12m^2 \phi^a\phi_a  + m\bar{\chi}^{\dot{a}}\chi_{\dot{a}}\;.
 \eea
This is invariant under ${\cal N}=4$ supersymmetry,
 \bea
   \delta_{\epsilon}\phi^{a} \ = \
   \bar{\epsilon}^{a\dot{a}}\chi_{\dot{a}}\;, \qquad
   \delta_{\epsilon}\chi_{\dot{a}} \ = \
   \tfrac{i}{2}\gamma^{\mu}\partial_{\mu}\phi^{a}\epsilon_{a\dot{a}}
   +\tfrac{1}{2}m\phi^{a}\epsilon_{a\dot{a}}\;,
 \eea
where the transformation parameters $\epsilon_{a\dot{a}}$ satisfy
the reality constraint
 \bea\label{real}
   \epsilon^{a\dot{a}} \ \equiv \ (\epsilon_{a\dot{a}})^{*} \ = \
   -\varepsilon^{ab}\varepsilon^{\dot{a}\dot{b}}\epsilon_{b\dot{b}}\;.
 \eea
Though this theory looks quite conventional and manifestly preserves
the R-symmetry, the latter actually acts as non-central charges in
order to prevent states beyond spin-1/2. To verify this, we compute
the commutator of the supersymmetry transformations. On the scalars
one finds
 \bea
   \big[ \delta_{\epsilon_1},\delta_{\epsilon_2}\big ] \phi^{a}
   \ = \ \xi^{\mu}\partial_{\mu}\phi^{a} +
   m\delta_{\Lambda}^{L}\phi^{a}\;,
 \eea
where apart from the usual translations parameterized by
 \bea
   \xi^{\mu} \ = \
   \tfrac{i}{4}\left(\bar{\epsilon}_{1\;a\dot{a}}\gamma^{\mu}\epsilon_{2}{}^{a\dot{a}}
   -\bar{\epsilon}_{2\;a\dot{a}}\gamma^{\mu}\epsilon_{1}{}^{a\dot{a}}\right)\;,
 \eea
the $\delta_{\Lambda}^{L}$ denotes an $\text{SU}(2)_L$ R-symmetry
transformation with parameters
 \bea
   \Lambda^{ab} \ = \
   \tfrac{1}{2}\varepsilon^{bc}\left(\bar{\epsilon}_{2}{}^{a\dot{a}}
   \epsilon_{1\;c\dot{a}}-\bar{\epsilon}_{1}{}^{a\dot{a}}
   \epsilon_{2\;c\dot{a}}\right)\;.
 \eea
Note that the symmetry of $\Lambda^{ab}$ is ensured by the reality
condition (\ref{real}). Similarly, one derives for the fermions
closure into translations and non-central terms corresponding to
$\text{SU}(2)_R$, up to the fermionic equations of motion.

\subsection{${\cal N}=8$}
Let us now discuss the scalar multiplets with ${\cal N}=8$
supersymmetry. The massive multiplet consists of eight real scalars
and eight real Majorana spinors, $(\phi^{A},\chi^{\dot{A}})$, the
former being in the spinor representation of $\text{SO}(8)$ and the
latter in the conjugate spinor representation. (Due to
$\text{SO}(8)$ triality this assignment of representations is rather
arbitrary.) The simplest case of a free massive Lagrangian is given
by
 \bea\label{N=8action}
   {\cal L} \ = \
   \ft12\partial^{\mu}\phi^{A}\partial_{\mu}\phi^{A}-i\bar{\chi}^{\dot{A}}\gamma^{\mu}\partial_{\mu}
   \chi^{\dot{A}}-\ft12 m^2\phi^{A}\phi^{A}-m\bar{\Gamma}^{1234}_{\dot{A}\dot{B}}\bar{\chi}^{\dot{A}}\chi^{\dot{B}}\;.
 \eea
Here we have restricted to one multiplet (otherwise $\phi$ and
$\chi$ would carry an additional $\text{SO}(N)$ index labeling the
multiplets), and ignored possible gauge couplings as in the massive
deformation of multiple M2-branes
\cite{Gomis:2008cv,Hosomichi:2008qk}. The supersymmetry parameter
transforms as a vector under $\text{SO}(8)$, and we have the
following supersymmetry rules
 \bea\label{N=8susy}
  \delta\phi^{A} \ = \
  \Gamma^{I}_{A\dot{A}}\bar{\epsilon}^{I}\chi^{\dot{A}}\;, \qquad
  \delta\chi^{\dot{A}} \ = \ \tfrac{i}{2}\gamma^{\mu}\partial_{\mu}\phi^{A}\Gamma^{I}_{A\dot{A}}
  \epsilon^{I}-\tfrac{1}{2}m\bar{\Gamma}^{1234}_{\dot{A}\dot{B}}\Gamma^{I}_{A\dot{B}}\phi^{A}\epsilon^{I}\;.
 \eea
The $\text{SO}(8)$ symmetry is explicitly broken to
$\text{SO}(4)\times \text{SO}(4)$ due to the presence of the
$\bar{\Gamma}^{1234}$ matrix in the mass term. This matrix has an
equal number of positive and negative eigenvalues and hence the
theory is parity-even. Let us mention that this theory can be
derived from gauged supergravity (see the discussion below,
\cite{Bergshoeff:2008ix}), in which the embedding tensor satisfies a
self-duality constraint in agreement with the fact that the
multiplets above require a fixed factor between the two
$\text{SO}(4)$ contributions.

We finally should comment on the following peculiarity. As far as
invariance of the action is concerned, the mass matrix
$\bar{\Gamma}^{1234}_{\dot{A}\dot{B}}$ can equally be replaced by
the $\text{SO}(8)$ invariant $\delta_{\dot{A}\dot{B}}$ in
(\ref{N=8action}). The analogue of the supersymmetry transformations
(\ref{N=8susy}) then close into $\text{SO}(8)$ rotations. In fact,
this free action has an $\text{SO}(8)\times \text{SO}(8)$ symmetry,
with the first factor acting on the bosons and the second factor
acting on the fermions. However, the presence of two independent
$\text{SO}(8)$ groups violates covariance of the supersymmetry
variations, due to the fact that $\Gamma^{I}_{A\dot{A}}$ is an
invariant tensor only with respect to a single $\text{SO}(8)$.
Consequently, these supersymmetry transformations will not close
with the $\text{SO}(8)$ generators. Rather, they will close into a
sort of generalized supersymmetry, in which instead of the
combination $\Gamma^{I}_{A\dot{A}}\epsilon^{I}$ a set of $64$
independent parameters $\epsilon_{A\dot{A}}$ appear. This is indeed
a symmetry which, however, is clearly an artefact of the free
theory. Moreover, these `supersymmetry' transformations will not
close into ordinary translations. This is not what we want for a
supersymmetric theory, in particular, it will not be extendable to
an interacting theory --- in contrast to (\ref{N=8action}). Thus we
will not consider this possibility any further.

\section{The ${\cal N}=16$ massive gauge theory}

In this section we construct the topologically massive gauge theory
announced in the introduction. We construct the theory by taking the
limit of gauged ${\cal N}=16$ supergravity to global supersymmetry
by decoupling gravity, following
\cite{Bergshoeff:2008ix,Bergshoeff:2008bh}. In order to illustrate
the procedure we will first in sec.~4.1 perform the limit of
ungauged supergravity, which results in a massless conformally
invariant theory, and then explain the limit for gauged supergravity
in sec.~4.2. The final result for the topologically massive
deformation is presented  in sec.~\ref{result}.

\subsection{The ${\cal N} = 16$ massless theory}

Ungauged ${\cal N}=16$ supergravity has been constructed in
\cite{Marcus:1983hb}, to which we refer the reader for further
details. The field content consists of $128$ scalar fields
$\phi^{A}$, parameterizing the coset space
$\text{E}_{8(8)}/\text{SO}(16)$, and $128$ Majorana fermions
$\chi^{\dot{A}}$.\footnote{The indices $I,J=1,\ldots,16$,
$A=1,\ldots, 128$ and $\dot{A}=1,\ldots,128$ refer now to the
vector, spinor and conjugate spinor representation of
$\text{SO}(16)$.} The metric $e_{\mu}{}^{a}$ and the $16$ gravitini
$\psi_{\mu}^{I}$ are purely topological in three dimensions and thus
do not add any propagating degrees of freedom. The Lagrangian is
given by \cite{Marcus:1983hb}
 \bea\label{ungauged}
  {\cal L} \ = \
  -\frac{1}{4\kappa^2}eR+\frac{1}{2}\varepsilon^{\mu\nu\rho}\bar{\psi}_{\mu}^{I}D_{\nu}\psi_{\rho}^{I}
  +\frac{1}{4\kappa^2}eg^{\mu\nu}P_{\mu}{}^{A}P_{\nu}{}^{A}
  -\frac{i}{2}e\bar{\chi}^{\dot{A}}\gamma^{\mu}D_{\mu}\chi^{\dot{A}}+\cdots\;,
 \eea
where we ignored higher-order terms, as these will drop out upon
taking the limit to global supersymmetry. Here, $P_{\mu}{}^{A}$ is
the non-compact part of the Maurer-Cartan forms defined in terms of
the $\text{E}_{8(8)}$--valued group element ${\cal V}(x)$ as
 \bea
  {\cal V}^{-1}\partial_{\mu}{\cal V} \ = \
  P_{\mu}{}^{A}Y^{A}+\ft12{\cal Q}_{\mu}{}^{IJ}X^{IJ}\;,
 \eea
where $t^{\cM}\equiv (X^{IJ},Y^{A})$ with adjoint indices
$\cM,\cN,\ldots=1,\ldots,248$ are the $\frak{e}_{8(8)}$ generators
in the $\text{SO}(16)$ decomposition ${\bf 248}\rightarrow {\bf 120}+{\bf
128}$. Upon gauge fixing the local $\text{SO}(16)$ symmetry, the
group--valued ${\cal V}$ can be parameterized in terms of the scalar
fields as ${\cal V}(x) = \exp\left(\phi^{A}(x)Y^{A}\right)$, which
implies
 \bea
  P_{\mu}{}^{A} \ = \ \partial_{\mu}\phi^{A}+{\cal O}(\phi^2)\;.
 \eea
We finally note that we have kept the explicit dependence on
Newton's constant $\kappa$, which is of mass dimension $-\ft12$.
The dimensions of the fields are
$(h_{\mu\nu},\psi_{\mu}^{A},\chi^{\dot{A}},\phi^{A})=(\ft12,1,1,0)$,
with $h_{\mu\nu}$ denoting the fluctuations of the metric around
Minkowski space, $g_{\mu\nu}=\eta_{\mu\nu}+\kappa h_{\mu\nu}$.

Let us now decouple gravity by sending $\kappa\rightarrow 0$. In
order for this limit to be non-singular, we need to rescale the
scalar fields as $\phi^{A}\rightarrow \kappa \phi^{A}$
\cite{Bergshoeff:2008ix}. After setting the topological supergravity
multiplet $(h_{\mu\nu},\psi_{\mu}^{I})$ to zero, the resulting
action describes the free theory
 \bea\label{freeaction}
  {\cal L}_{0} \ = \
  \ft14\partial^{\mu}\phi^{A}\partial_{\mu}\phi^{A}-\tfrac{i}{2}\bar{\chi}^{\dot{A}}
  \gamma^{\mu}\partial_{\mu}\chi^{\dot{A}}\;,
 \eea
while the supersymmetry transformations of \cite{Marcus:1983hb}
reduce to
 \bea\label{freeSUSY}
  \delta_{\epsilon}\phi^{A} \ = \
  \Gamma^{I}_{A\dot{A}}\bar{\chi}^{\dot{A}}\epsilon^{I}\;, \qquad
  \delta_{\epsilon}\chi^{\dot{A}} \ = \
  \tfrac{i}{2}\Gamma^{I}_{A\dot{A}}\gamma^{\mu}\partial_{\mu}\phi^{A}\epsilon^{I}\;.
 \eea
One may easily convince oneself that (\ref{freeSUSY}) leaves
(\ref{freeaction}) invariant, i.e., in spite of the fact that ${\cal
N}=16$ represents $32$ real supercharges, it is a symmetry of the
globally supersymmetric action (\ref{freeaction}). As we noted in
the introduction, the existence of this theory is not in conflict
with the `higher-spin barrier', which in dimensions $D\geq 4$
excludes globally supersymmetric theories with more than $16$
supercharges. In fact, it has already been noticed in
\cite{Marcus:1983hb} that free supersymmetric theories in $D=3$ can
be written for any ${\cal N}=8k$. One simply uses the fact that for
multiples of $8$, $\text{SO}({\cal N})$ possesses two inequivalent real
spinor representations of the same dimension, with invariant tensor
$\Gamma^{I}_{A\dot{A}}$, such that (\ref{freeSUSY}) immediately
extends to ${\cal N}=8k$.

\subsection{The ${\cal N}=16$ massive theory}
We now turn to the global limit of gauged supergravity, which will
lead to a massive deformation of (\ref{freeaction}), featuring in
addition to massive scalars and spinors topologically massive gauge vectors. The latter is in
agreement with the general structure of BPS multiplets discussed
in the previous section.

The gauged ${\cal N}=16$ supergravity as constructed in
\cite{Nicolai:2000sc,Nicolai:2001sv} is completely determined by
means of the so-called embedding tensor
$\Theta_{\cM\cN}=\Theta_{\cN\cM}$. The latter encodes the subgroup
of the rigid invariance group $\text{E}_{8(8)}$ that is gauged by
determining the covariant derivatives
 \bea\label{cov}
  D_{\mu} \ = \
  \partial_{\mu}+\Theta_{\cM\cN}A_{\mu}{}^{\cM}t^{\cN}\;.
 \eea
More precisely, one introduces $248$ vector fields in order to
perform the gauging which, however, will only enter through a
topological Chern-Simons term and as such do not alter the counting
of degrees of freedom. The action is given by
 \begin{eqnarray}\label{gauged}
  {\cal L} &=& \cL_0
  -\frac{1}{4}\varepsilon^{\mu\nu\rho}A_{\mu}{}^{\cM}\Theta_{\cM\cN}\left(\partial_{\nu}A_{\rho}{}^{\cN}
  +\frac{1}{3}\Theta_{\cK\cS}f^{\cN\cS}{}_{\cL}A_{\nu}{}^{\cK}A_{\rho}{}^{\cL}\right)\\
  \nonumber
  &&+\frac{1}{2\kappa^2}eA_{1}^{IJ}\bar{\psi}_{\mu}^{I}\gamma^{\mu\nu}\psi_{\nu}^{J}
  +\frac{i}{\kappa^2}eA_2^{I\dot{A}}\bar{\chi}^{\dot{A}}\gamma^{\mu}\psi_{\mu}^{I}
  +\frac{1}{2\kappa^2}eA_3^{\dot{A}\dot{B}}\bar{\chi}^{\dot{A}}\chi^{\dot{B}}-\frac{1}{\kappa^6}eV\;.
 \end{eqnarray}
Here, $\cL_0$ denotes the ungauged Lagrangian (\ref{ungauged}), in
which all derivatives have been replaced by the covariant
derivatives (\ref{cov}). The scalar-dependent Yukawa couplings
parameterized by $A_{1,2,3}$ and the scalar potential $V$, which can
be written as a square in $A_1$ and $A_2$, are completely determined
by the embedding tensor. Their expressions can be found in \cite{Nicolai:2000sc,Nicolai:2001sv}.
The action \eqref{gauged} is invariant under local
supersymmetry, provided the fermionic variations acquire shift terms
proportional to $\Theta_{\cM\cN}$,
 \bea
  \delta\psi_{\mu}^{I} \ = \ \delta_0\psi_{\mu}^{I}
  +iA_{1}^{IJ}\gamma_{\mu}\epsilon^{J}\;,
  \qquad
  \delta\chi^{\dot{A}}=\delta_0\chi^{\dot{A}}+A_2^{I\dot{A}}\epsilon^{I}\;,
 \eea
and provided the embedding tensor satisfies a linear and quadratic
constraint. The explicit form of the linear constraint is given by
eq.~(4.6) of ref.~\cite{Nicolai:2001sv}. The quadratic constraint
follows by requiring gauge invariance of (\ref{gauged}) and,
consequently, invariance of the embedding tensor. It reads
 \bea\label{quadconstr}
  \Theta_{\cP\cK}\Theta_{\cL(\cM}f^{\cK\cL}{}_{\cN)} \ = \ 0 \;,
 \eea
where $f$ denotes the $\text{E}_{8(8)}$ structure constants.

Let us now discuss the decoupling limit $\kappa\rightarrow 0$.
Splitting the $\text{E}_{8(8)}$ indices under $\text{SO}(16)$,
$\cM=([IJ],A)$, we obtain three components of the embedding tensor,
$\Theta_{IJ,KL}$, $\Theta_{IJ,A}$, $\Theta_{AB}$, and
correspondingly two types of gauge fields, $A_{\mu}{}^{IJ}$ and
$A_{\mu}{}^{A}$. As was shown in \cite{Bergshoeff:2008ix}, this
limit is only non-singular and admits non-trivial supersymmetry
transformations for the gauge vectors, provided one performs first
certain rescalings with $\kappa$. More precisely, the components of
the embedding tensor need to be rescaled with $\kappa^{2}$ and the
gauge vectors by $\kappa^{-1}$. Afterwards, the $\text{SO}(16)$
gauge vectors have to be set to zero, as these belong to the
supergravity multiplet. This is in accordance with the fact that in
globally supersymmetric theories the R-symmetry group cannot be
gauged. Instead, the components of $\Theta$ in the $\text{SO}(16)$
direction will give rise to massive deformations, as we will see
below.

The condition of a non-singular limit requires moreover that certain
components of the embedding tensor are set to zero, or in other
words, that there are additional linear constraints. These can be
determined by expanding the tensors $A_{1,2,3}$ in powers of the
scalar fields and $\Theta$ and inspecting their scaling behavior
with $\kappa$, as has been shown in \cite{Bergshoeff:2008ix}. Rather
than repeating these steps in detail here, we will just state the
results and refer the reader to \cite{Bergshoeff:2008ix} and
\cite{Nicolai:2001sv} for explicit formulae. In total, one finds
that the available components of the embedding tensor are
$\Theta_{IJ,KL}$, satisfying $\Theta_{IK,JK}=0$, and $\Theta_{AB}$.
Together with the linear constraints of \cite{Nicolai:2001sv} this
in turn implies
 \bea\label{gensol}
 \begin{split}
  \Theta_{IJ,KL} \ &= \ f_{IJKL} \ \equiv \ f_{[IJKL]}\;, \qquad
  \Theta_{IJ,A} \ = \ 0\;, \\
  \Theta_{AB} \ &= \ \tfrac{1}{96}\Gamma^{IJKL}_{AB}f_{IJKL}\;,
 \end{split}
 \eea
i.e., the embedding tensor is parameterized in terms of a totally
antisymmetric 4-rank tensor $f_{IJKL}$. Without referring further to
the supergravity limit we will present the Lagrangian and
supersymmetry rules of the ${\cal N}=16$ massive gauge theory in the
following subsection.

\subsection{${\cal N}=16$ action and supersymmetry transformations}\label{result}
We find that the Lagrangian corresponding to the action of the ${\cal N}=16$ massive
gauge theory is given by
 \begin{eqnarray}\label{32massive}
  \cL  &=& \frac{1}{4}
  D^{\mu}\phi^{A}D_{\mu}\phi^{A}-\frac{i}{2}\bar{\chi}^{\dot{A}}\gamma^{\mu}\partial_{\mu}\chi^{\dot{A}}
  -\frac{1}{4}\varepsilon^{\mu\nu\rho}\Theta_{AB}A_{\mu}{}^{A}\partial_{\nu}A_{\rho}{}^{B}
  \\ \nonumber
  &&+\frac{1}{96}\bar{\Gamma}^{IJKL}_{\dot{A}\dot{B}}\Theta_{IJ,KL}\bar{\chi}^{\dot{A}}\chi^{\dot{B}}
  -\frac{1}{16}A_2^{I\dot{A}}A_2^{I\dot{A}}\;,
 \end{eqnarray}
where we defined
 \bea
 \begin{split}
  D_{\mu}\phi^{A} \ &= \
  \partial_{\mu}\phi^{A}+\Theta_{AB}A_{\mu}{}^{B}\;, \\
  A_2^{I\dot{A}} \ &= \
  \tfrac{1}{16}\left(\Gamma^{JKL}_{A\dot{A}}\Theta_{IJ,KL}+\tfrac{1}{12}\Gamma^{IJKLM}_{A\dot{A}}\Theta_{JK,LM}\right)
  \phi^{A}\;.
 \end{split}
 \eea
The ${\cal N}=16$ supersymmetry transformations (corresponding to $32$ real
supercharges) read
 \begin{eqnarray}\label{32SUSY}\nonumber
  \delta_{\epsilon}\phi^{A} &=&
  \Gamma^{I}_{A\dot{A}}\bar{\chi}^{\dot{A}}\epsilon^{I}\;, \\
  \delta_{\epsilon}\chi^{\dot{A}} &=&
  \tfrac{i}{2}\Gamma^{I}_{A\dot{A}}\gamma^{\mu}D_{\mu}\phi^{A}\epsilon^{I}+A_2^{I\dot{A}}\epsilon^{I}\;,\\
  \nonumber
  \delta_{\epsilon}A_{\mu}{}^{A} &=&
  i\Gamma^{I}_{A\dot{A}}\bar{\epsilon}^{I}\gamma_{\mu}\chi^{\dot{A}}\;.
 \end{eqnarray}
In the global limit there is a remnant of the quadratic constraint
(\ref{quadconstr}), which reads
 \bea\label{globalquad}
  \Gamma^{IJ}_{CD}\Theta^{}_{AC}\Theta^{}_{BD}+\ft12\Gamma^{KL}_{AC}\Theta^{}_{CB}\Theta^{}_{IJ,KL}
  \ = \ 0 \;.
 \eea
To summarize, the action corresponding to (\ref{32massive}) is
invariant under the ${\cal N}=16$ supersymmetry variations
(\ref{32SUSY}), provided the components of the embedding tensor are
given by (\ref{gensol}), satisfying the quadratic constraint
(\ref{globalquad}).

Let us now determine the physical content of (\ref{32massive}). The
scalar potential quadratic in $A_2$ reduces to pure mass terms for
$\phi^{A}$. Similarly, the Yukawa couplings involving
$\chi^{\dot{A}}$ lead to mass terms for the spinors. To determine
the number of massive spin-0 and spin-1 degrees of freedom, we note
that by virtue of (\ref{globalquad}) the Lagrangian
\eqref{32massive} is invariant under the local shift symmetry
 \bea\label{shiftsym}
  \delta_{\Lambda}\phi^{A} \ = \ \Theta_{AB}\Lambda^{B}\;, \qquad
  \delta_{\Lambda}A_{\mu}{}^{A} \ = \ -\partial_{\mu}\Lambda^{A}\;.
 \eea
Therefore, the scalar potential does not depend on all scalar
fields, but only on a subset determined by the choice of embedding
tensor, which are precisely those that become massive due to the
presence of $A_2$. The complementary scalar fields can in turn be
gauged to zero by virtue of (\ref{shiftsym}). The field equations
for the corresponding vector fields then take the form of massive
self-duality equations,
 \bea
  \Theta_{AB}\left(F_{\mu\nu}{}^{B}-\Theta_{BC}\varepsilon_{\mu\nu\rho}A^{\rho\;C}\right)
  \ = \ 0\;.
 \eea
After acting with $\partial^{\mu}$, one obtains the standard massive
Yang-Mills equation with mass matrix $\Theta_{AB}$
\cite{Townsend:1983xs}. In other words, the vectors corresponding to
a zero eigenvalue of $\Theta_{AB}$ disappear from the Lagrangian,
leaving a massive scalar, while a non-zero eigenvalue indicates a
massive spin-1 field in a St\"uckelberg formulation.

According to the results summarized in table~2, the $128$ bosonic
degrees of freedom should be distributed, for any choice of
embedding tensor, among $96$ massive spin-0 scalars and $32$ massive
spin-1 vectors. In fact, the possible solutions of
(\ref{globalquad}) are quite restricted. It turns out that a
solution is given by various copies of the $\text{SO}(4)$
Levi-Civita symbol. Thus the R-symmetry group $\text{SO}(16)$ is
broken into $\text{SO}(4)\times \cdots\times \text{SO}(4)$.
Splitting the $\text{SO}(16)$ indices into four blocks of
$\text{SO}(4)$ vector indices $i,j,\ldots=1,\ldots,4$, the solution
is given by
 \bea\label{thetasol}
   \Theta_{ij,kl} \ = \ f_{ijkl} \ = \ m\;\varepsilon_{ijkl}\;, \qquad {\rm etc.}
 \eea
Moreover, the parameter $m$ is restricted by (\ref{globalquad}) to
be the same for all four copies of $\text{SO}(4)$.\footnote{To be
precise, there is slightly more freedom in that certain relative
signs between the four sectors are not fixed. However, one may check
that these choices lead to the same mass matrices in
(\ref{32massive}) and so do not represent physically different
theories.} This is in agreement with the analysis of the foregoing
section, since in the commutator algebra of the supersymmetry
transformations (\ref{32SUSY}), these parameters multiply the
non-central $\text{SO}(4)$ generators, which on the other hand were
required to be equal in order to have the maximal multiplet
shortening. For the same reason we do not expect the existence of
any other solutions of (\ref{globalquad}). We verified with {\tt
Mathematica} that inserting (\ref{thetasol}) into (\ref{gensol})
gives rise to the correct number of zero eigenvalues of
$\Theta_{AB}$, in agreement with the expected number of massive
spin-$0$ and spin-$1$ degrees of freedom (including negative and
positive helicities). Moreover, also the scalar mass matrix
determined by $A_2^{I\dot{A}}$ and the fermionic mass matrix give
rise to the expected eigenvalues.

Let us finally comment on the full gauged supergravity, which gives
rise to the given Poincar\'e invariant theory upon decoupling
gravity.\footnote{We would like to thank Henning Samtleben for
discussions on this point and for bringing
ref.~\cite{Fischbacher:2002fx} to our attention.} This has to be the
$\text{SO}(4,4)\times \text{SO}(4,4)$ gauging analyzed in
\cite{Fischbacher:2002fx}, since it has a unique Minkowski ground
state, whose mass spectrum coincides with the spectrum above. It
would be instructive to study the precise embedding in more detail,
but we will leave this for future work.

\subsection{Interacting theories beyond ${\cal N}=8$?}
One may wonder whether the limit of gauged supergravity allows the
construction of interacting globally supersymmetric theories beyond
${\cal N}=8$. First of all, the free massive deformations as for the
$\cN=16$ case just described will also exist for ${\cal N}=9,10,12$,
simply by taking an embedding tensor in the R-symmetry direction.
Concerning the problem of a limit which leaves an interacting
theory, $\cN=12$ seems to be a promising candidate, since in this
case the coset space in supergravity is
$\text{E}_{7(-5)}/(\text{SO}(12)\times \text{SU}(2))$. In
particular, the local subgroup $H$ consists not only of the
R-symmetry group $\text{SO}(12)$, but also of the non-abelian
complement $\text{SU}(2)$. If a gauging only of this $\text{SU}(2)$
is possible, this would give rise in the limit to a conformally
invariant $\text{SU}(2)$ Chern-Simons theory. Unfortunately, the
general solution of the constraints for compact gauge groups given
in \cite{deWit:2003ja} (see their eq.~(5.17)) does not allow to
consistently switch off the gaugings in the R-symmetry direction.
Since, as shown in \cite{Bergshoeff:2008ix}, the components of the
embedding tensor in the R-symmetry (${\rm SO}(12)$) and global
symmetry (${\rm SU}(2)$) directions scale differently with Newton's
constant, it follows that these gaugings do not allow a consistent
flat space limit. Thus we conclude that $\cN=12$ supergravity does
not give rise to a non-abelian, interacting theory.

\section{Discussion and Outlook}
In this paper we analyzed an extension of Poincar\'e supersymmetry
in three dimensions by non-central R-symmetry generators, both at
the level of the supermultiplets and at the level of field
theoretical realizations. We found an unconventional type of
multiplet shortening, which goes beyond the standard one known from
central charges and BPS multiplets. In particular, the usual bounds
for supersymmetry are stretched by a factor of 2 in that scalar
multiplets with maximum spin $\ft12$ are possible up to 16
supercharges and vector multiplets with spin $1$ up to 32
supercharges. For the latter we determined a field theoretical
realization with topologically massive gauge fields by decoupling
gravity from gauged ${\cal N}=16$ supergravity.

This unexpected phenomenon suggests interesting further research.
First of all, the  massive ${\cal N}=16$ theory we constructed in
this paper is a free theory. Since the discussed mechanism of
multiplet shortening happens also for interacting theories (as the
massive deformations of the Bagger-Lambert theory), the question
arises whether interacting theories with $\cN>8$ exist. One approach
to derive more general theories might be to take the limit of
supergravity to non-flat backgrounds. Furthermore, it would be
interesting to find out whether the given model or extensions
thereof has a direct physical interpretation, say in the context of
brane dynamics. Perhaps
 supersymmetry enhancement as in
\cite{Itzhaki:2005tu}  plays a role here.

Finally, let us note that requiring maximal spin 2 allows
supersymmetry up to ${\cal N}=32$ corresponding to 64 supercharges,
and so one may hope to construct supergravity theories with this
amount of supersymmetry. Actions for \textit{massive}
propagating spin-2 fields do exist in three dimensions. Here, the
usual (topological) Einstein-Hilbert action is extended by a
gravitational Chern-Simons term, quite analogous to the topological
mechanism for spin-1 fields encountered above \cite{Deser:1981wh}.
In fact, these can even be made supersymmetric \cite{Deser:1982sw}.
However, the supermultiplets discussed in sec.~2 require that the
spin-2 states transform non-trivially under the R-symmetry group, or
in other words, this would require a multi-graviton theory. While
theories of this type are usually considered to be consistent only
in case of an infinite number of spin-2 fields \cite{Hohm:2005sc},
it might be worth investigating whether this unconventional
framework allows for new possibilities.

\subsection*{Acknowledgments}

We thank B.~de Wit, J.~Engquist, H.~Lin for useful discussions and
especially H.~Nicolai for asking the right question. This work was
partially supported by the European Commission FP6 program
MRTN-CT-2004-005104EU and by the INTAS Project 1000008-7928.

\begin{appendix}
\renewcommand{\theequation}{\Alph{section}.\arabic{equation}}

\section{Conventions and useful relations} \setcounter{equation}{0}

\subsection{Spinor conventions in $D=3$}\label{3Dspinor}
For the $\text{SO}(1,2)$ gamma matrices we choose the purely imaginary
basis
 \bea
  \gamma^0 \ = \ \sigma_2\;, \qquad
  \gamma^1 \ = \ i\sigma^3\;, \qquad
  \gamma^2 \ = \ i\sigma^1\;,
 \eea
which satisfies the Clifford algebra
$\{\gamma^{\mu},\gamma^{\nu}\}=2\eta^{\mu\nu}$ for
$\eta^{\mu\nu}=(+--)$. The charge conjugation matrix is defined as
 \bea
  C \ = \ \gamma^0\;,
 \eea
satisfying
 \bea
  C(\gamma^{\mu})^{T}C^{-1} \ = \ -\gamma^{\mu}\;.
 \eea
Consequently, the matrices $(\gamma^{\mu}C)_{\alpha\beta}$ are symmetric in the
spinor indices $\alpha,\beta$. We define the bilinear
$\bar{\psi}\psi$ through
 \bea\label{defbar}
   \bar{\psi} \ = \ \psi^{T}\gamma^{0}\;,
 \eea
which is invariant under the real three-dimensional Lorentz group
$SL(2,\mathbb{R})$. In particular, it can be defined without complex
conjugation, even if the spinors are not real. The identities
 \bea
   \bar{\psi}\chi \ = \ \bar{\chi}\psi\;, \qquad
   \bar{\psi}\gamma^{\mu}\chi \ = \ -\bar{\chi}\gamma^{\mu}\psi\;,
   \qquad {\rm etc.}
 \eea
readily follow. One can impose a Majorana condition on a spinor
$\psi$, which reads $\bar{\psi^{*}} = \psi^{T}C$. In the given
basis, this means that $\psi$ is real. For Majorana spinors $\psi$,
the bilinear $\bar{\psi}\psi$ is real by virtue of the convention
$(\psi_1\psi_2)^{*}=\psi_2^{*}\psi_1^{*}$. We note that due to the
definition (\ref{defbar}), there are two different real Lorentz
invariant bilinears for complex Dirac spinors $\chi_{\dot{a}}$ (as,
e.g., used for ${\cal N}=4$ in the main text), namely
$\bar{\chi}^{\dot{a}}\chi_{\dot{a}}$ and
$\bar{\chi}_{\dot{a}}\chi_{\dot{b}}+{\rm h.c.}$.

\subsection{$\text{SO}(4)$ conventions}\label{so4app}
The $\text{SO}(4)$ generators are given by $M_{ij}=-M_{ji}$,
$i,j,\ldots=1,\ldots,4$, satisfying the algebra (\ref{so4}). In
order to exhibit the isomorphism $\text{SO}(4)\cong \text{SU}(2)_{L}\times
\text{SU}(2)_{R}$ it is convenient to introduce spinor indices
$a,b,\ldots=1,2$, $\dot{a},\dot{b},\ldots=1,2$ and to relate $\text{SO}(4)$
vector indices to bispinors via $\Gamma^{i}_{a\dot{a}}\equiv (i{\bf
1},\sigma^{1},\sigma^{2},\sigma^{3})$. This allows to introduce
generators
 \bea
  M_{ab} \ = \ -\ft14\Gamma^{ij}_{ab}M_{ij}\;, \qquad
  M_{\dot{a}\dot{b}} \ = \
  -\ft14\bar{\Gamma}_{\dot{a}\dot{b}}^{ij}M_{ij}\;,
 \eea
or, inversely,
 \bea
  M^{ij} \ = \
  \ft12\left(\Gamma^{ij}_{ab}M^{ab}+\bar{\Gamma}_{\dot{a}\dot{b}}^{ij}M^{\dot{a}\dot{b}}\right)\;,
 \eea
which are both symmetric in their respective spinor indices. Here we
have defined
 \bea
  \begin{split}
   \Gamma^{ij}_{ab} \ &= \
   \ft12\varepsilon^{\dot{a}\dot{b}}\left(\Gamma^{i}_{a\dot{a}}\Gamma^{j}_{b\dot{b}}-
   \Gamma^{j}_{a\dot{a}}\Gamma^{i}_{b\dot{b}}\right)\;,
   \\
   \bar{\Gamma}^{ij}_{\dot{a}\dot{b}} \ &= \
   \ft12\varepsilon^{ab}\left(\Gamma^{i}_{a\dot{a}}\Gamma^{j}_{b\dot{b}}-
   \Gamma^{j}_{a\dot{a}}\Gamma^{i}_{b\dot{b}}\right)\;,
  \end{split}
 \eea
where we introduced the $\text{SU}(2)$ invariant Levi-Civita symbol
$\varepsilon^{ab}$ (with $\varepsilon^{12}=\varepsilon_{12}=+1$),
which allows to raise and lower indices and which satisfies
$\varepsilon_{ac}\varepsilon^{cb}=-\delta_{a}{}^{b}$. In terms of
$M_{ab}$ and $M_{\dot{a}\dot{b}}$ the $\text{SO}(4)$ algebra (\ref{so4})
takes explicitly the direct product form $\text{SU}(2)_{L}\times
\text{SU}(2)_{R}$,
 \bea
  \begin{split}
   \big[ M_{ab},M_{cd} \big] \ &= \
   \ft12\left(\varepsilon_{ac}M_{bd}+\varepsilon_{bc}M_{ad}
   +\varepsilon_{ad}M_{bc}+\varepsilon_{bd}M_{ac}\right)\;, \\
   \big[ M_{\dot{a}\dot{b}},M_{\dot{c}\dot{d}} \big] \ &= \
   \ft12\left(\varepsilon_{\dot{a}\dot{c}}M_{\dot{b}\dot{d}}+\varepsilon_{\dot{b}\dot{c}}M_{\dot{a}\dot{d}}
   +\varepsilon_{\dot{a}\dot{d}}M_{\dot{b}\dot{c}}+\varepsilon_{\dot{b}\dot{d}}M_{\dot{a}\dot{c}}\right)\;,
   \\ \big[ M_{ab}, M_{\dot{a}\dot{b}} \big] \ &= \ 0\;.
  \end{split}
 \eea
Moreover, in this language the raising and lowering operators
introduced in the main text are bispinors $a^{\dagger}_{a\dot{a}}$,
etc., and satisfy
 \bea\label{semiraising}
  \big[ M_{ab},a^{\dagger}_{c\dot{c}} \big] \ = \
  \ft12\left(\varepsilon_{ac}a^{\dagger}_{b\dot{c}}+\varepsilon_{bc}a^{\dagger}_{a\dot{c}}\right)\;,
  \qquad
  \big[ M_{\dot{a}\dot{b}},a^{\dagger}_{c\dot{c}} \big] \ = \
  \ft12\left(\varepsilon^{}_{\dot{a}\dot{c}}a^{\dagger}_{c\dot{b}}+\varepsilon_{\dot{b}\dot{c}}a^{\dagger}_{c\dot{a}}\right)\;,
 \eea
and similarly for lowering operators.

The given basis is convenient in order to develop the representation
theory, since the generators immediately represent lowering and
raising operators for $\text{SU}(2)$. To see this, we split the
indices according to $a=(+,-)$ and $\dot{a}=(\dot{+},\dot{-})$. Then
one can identify the $\text{SU}(2)_{L}$ generators,
 \bea
  J_{+}^{L} \ = \ M_{++}\;, \qquad J_{-}^{L} \ = \ M_{--}\;, \qquad
  J_{3}^{L} \ = \ -M_{+-}\;,\
 \eea
satisfying the standard algebra
 \bea
  \big[ J_3^{L},J_{+}^{L} \big] \ = \ J_{+}^{L}\;, \qquad
  \big[ J_3^{L},J_{-}^{L} \big] \ = \ -J_{-}^{L}\;, \quad
  \big[ J_{+}^{L},J_{-}^{L}\big] \ = \ -2J_3\;,
 \eea
and analogously for $\text{SU}(2)_{R}$. This notation is chosen such that
the index structure of the raising and lowering operators directly
indicates how it increases (+) or decreases (--) the $\text{SU}(2)$ quantum
numbers of a given state. For instance, from (\ref{semiraising}) one
infers
 \bea
  \big[ J_{3}^{L},a^{\dagger}_{+\dot{-}}\big] \ = \ \ft12
  a^{\dagger}_{+\dot{-}}\;\;, \qquad
  \big[ J_{3}^{R},a^{\dagger}_{+\dot{-}}\big] \ = \ -\ft12
  a^{\dagger}_{+\dot{-}}\;\;.
 \eea
Consequently, $a^{\dagger}_{+\dot{-}}$ increases the $\text{SU}(2)_{L}$
quantum number and decreases the $\text{SU}(2)_{R}$ quantum number by
$\ft12$.

We finally give some identities, which we found useful for relating
$\text{SO}(4)$ to $\text{SU}(2)$ quantities. The $\Gamma^{i}_{a\dot{a}}$ satisfy
 \bea
  \begin{split}
   \varepsilon^{ab}\varepsilon^{\dot{a}\dot{b}}\Gamma^{i}_{a\dot{a}}\Gamma^{j}_{b\dot{b}}
   \ &= \ -2\delta^{ij}\;, \qquad\qquad
   \Gamma^{i}_{a\dot{a}}\Gamma^{i}_{b\dot{b}} \ = \
   -2\varepsilon_{ab}\varepsilon_{\dot{a}\dot{b}}\;, \\
   \Gamma^{ij}_{ab}\Gamma^{j}_{c\dot{c}} \ &= \ -2\varepsilon^{}_{c(a}\Gamma^{i}_{b)\dot{c}}
   \;, \qquad
   \bar{\Gamma}^{ij}_{\dot{a}\dot{b}}\Gamma^{j}_{c\dot{c}} \ = \
   -2\varepsilon^{}_{\dot{c}(\dot{a}}\bar{\Gamma}^{i}_{\dot{b})c}\;,
  \end{split}
 \eea
while the $\Gamma^{ij}$ and $\bar{\Gamma}^{ij}$ obey the (anti-)self
duality relations
 \bea\label{relsign}
   \Gamma^{ij}_{ab} \ = \ \ft12
   \varepsilon^{ijkl}\Gamma^{kl}_{ab}\;, \qquad
   \bar{\Gamma}^{ij}_{\dot{a}\dot{b}} \ = \
   -\ft12\varepsilon^{ijkl}\bar{\Gamma}^{kl}_{\dot{a}\dot{b}}\;.
 \eea
Finally, we have the reality constraints
 \bea
   (\Gamma_i^{\dagger})^{a\dot{a}} \ = \
   -\varepsilon^{ab}\varepsilon^{\dot{a}\dot{b}}(\Gamma_i)_{b\dot{b}}\;
   \qquad
  (\Gamma_{ij}^{\dagger})^{ab} \ = \
  \varepsilon^{ac}\varepsilon^{bd}(\Gamma_{ij})_{cd}\;,
 \eea
such that the anti-hermiticity $(M^{ij})^{\dagger}=-M^{ij}$ implies
for the $\text{SU}(2)_{L}$ generators
 \bea
  M^{ab} \ = \ (M_{ab})^{\dagger} \ = \ -\varepsilon^{ac}\varepsilon^{bd}M_{cd}\;,
 \eea
and analogously for $\text{SU}(2)_R$.

\end{appendix}

\end{document}